# Image reconstruction from few views by $\ell_0$-norm optimization

Yuli Sun, Jinxu Tao [*]

*Department of Electronic Engineering and Information Science, University of Science and Technology of China, Hefei, Anhui 230027, People's Republic of China*

***Abstract:*** In the medical computer tomography (CT) field, total variation (TV), which is the $\ell_1$-norm of the discrete gradient transform (DGT), is widely used as regularization based on the comprehensive sensing (CS) theory. To overcome the TV model's disadvantageous tendency of uniformly penalize the image gradient and over smooth the low-contrast structures, an iterative algorithm based on the $\ell_0$-norm optimization of the DGT is proposed. To rise to the challenges introduced by the $\ell_0$-norm DGT, the algorithm uses a pseudo-inverse transform of DGT and adapts an iterative hard thresholding (IHT) algorithm, whose convergence and effective efficiency have been theoretically proven. The simulation demonstrates our conclusions and indicates that the algorithm proposed in this paper can obviously improve the reconstruction quality.

## 1. Introduction

Image reconstruction from insufficient data is an important issue in computed tomography (CT). Insufficient data problems occur quite frequently because of practical constraints due to the imaging hardware, scanning geometry, or ionizing radiation exposure.[1] The insufficient data may arise from various forms, but in this work we only consider the few views reconstruction problem. Reconstruction by using standard analytic algorithms such as filtered back-projection (FBP) algorithm will lead to distortion and artifact.

Mathematically, image reconstruction from few views can be thought of as an underdetermined linear system. To overcome this ill-posed problem, we need making some constraint to narrow the solution space. One type of them is based on the compressive sensing (CS) theory proposed by Candes et al..[2] The main idea of CS is that most signals are sparse in some appropriate representations; that is, a majority of their coefficients are close or equal to zero, when represented in these proper domains.[3] Then, CS reconstructs these sparse signals by using the $\ell_1$-norm minimization.[3-7]

---

[*] Corresponding Author: Tel: +86-551-63601329; Email: tjingx@ustc.edu.cn.

Because the x-ray attenuation coefficient often varies mildly within an anatomical component, and large changes are usually confined around the borders of tissue structures, discrete gradient transform (DGT) has been widely utilized as a sparsifying operator in CS-inspired CT reconstruction.[5-14] Without loss of generality, the corresponding $\ell_1$-norm reconstruction problem can be described as

$$\min_{\mathbf{f}} \|\mathbf{f}\|_{TV} = \min_{\mathbf{f}} \|\nabla \mathbf{f}\|_1 \quad s.t. \quad \mathbf{Rf} = \mathbf{p}, \tag{1}$$

where $\mathbf{f}$ is the CT image to be reconstructed, $\mathbf{R}$ is the measurement matrix, $\mathbf{p}$ is the measured data and $\|\mathbf{f}\|_{TV}$ is the $\ell_1$-norm of the DGT image $\nabla \mathbf{f}$, which is also referred to as the well-known total variation (TV) of $\mathbf{f}$. However, the DGT does not satisfy the restricted isometry property (RIP) required by the CS theory, [3] the $\ell_1$-norm TV regularization is tending to uniformly penalize the image gradient and the low-contrast structures are sometimes over smoothed in the reconstruction.[14,15]

In this article, we focus on the $\ell_0$-norm of the DGT image $\nabla \mathbf{f}$, we consider the following $\ell_0$-norm minimization problem:

$$\mathbf{f} = \arg\min \|\nabla \mathbf{f}\|_0 \quad s.t. \quad \|\mathbf{Rf} - \mathbf{p}\|_2^2 \leq e. \tag{2}$$

The optimization problem in (2) causes the image estimation with a minimum $\ell_0$-norm constraint such that the $\ell_2$-distance between its projection data with the measured projection data within a given tolerance $e$.

To the best of our knowledge, there are seldom reconstruction algorithms using the $\ell_0$-norm of DGT. This is due to that there are two main challenges in solving (2): first, the $\ell_0$-norm minimization problem is known to be NP-hard; second, the DGT is not invertible in general, the Frobenius operator norm of its inverse grows linearly with the size of $\mathbf{f}$. In this work, we use the iterative hard thresholding (IHT) algorithm by Blumensath and Davies [16,17] to address the first challenge. We fix the sparsity of the DGT image to $S$, and consider this optimization problem:

$$\mathbf{f} = \arg\min \|\mathbf{R}\mathbf{f} - \mathbf{p}\|_2^2 \quad s.t. \quad \|\nabla \mathbf{f}\|_0 \leq S. \tag{3}$$

Inspired by the pseudo-inverse transform in the soft-threshold filtering by Yu and Wang,[7] we modify it and apply it for this problem to address the second challenge. The numerical simulation shows that the proposed $\ell_0$-norm algorithm can effectively improve the reconstruction quality.

The rest of the paper is structured as follows. In section 2, we briefly review the IHT algorithm and propose the new $\ell_0$-norm reconstruction algorithm. Section 3 presents the numerical results. In the end, we provide our conclusion in section 4.

## 2. Method

### 2.1. IHT algorithm

Blumensath and his collaborator proved the convergence of the IHT algorithm to solve the inverse problems regularized by a sparsity constrained.[16-20] Their major results can be summarized as follows.

A sparse or approximately sparse signal $\mathbf{f} = [f_1, f_2, \cdots, f_N] \in \mathbb{R}^N$ is sampled with a linear sampling operator $\mathbf{R} = (r_{m,n}) \in \mathbb{R}^M \times \mathbb{R}^N$, $M \ll N$. The samples $\mathbf{p} \in \mathbb{R}^M$ are potentially corrupted by observation noise $\mathbf{e} \in \mathbb{R}^M$, that is

$$\mathbf{p} = \mathbf{R}\mathbf{f} + \mathbf{e}. \tag{4}$$

Under the assumption that we know that $\mathbf{f}$ is $K$-sparse, then the reconstruction can be the following optimization problem:

$$\mathbf{f} = \arg\min \|\mathbf{R}\mathbf{f} - \mathbf{p}\|_2^2 \quad s.t. \quad \|\mathbf{f}\|_0 \leq K. \tag{5}$$

Initialized with $\mathbf{f}^0 = \mathbf{0}$, the IHT algorithm use the iteration[19-21]

$$\tilde{\mathbf{f}}^n = \mathbf{f}^{n-1} + \mu \mathbf{R}^T \left(\mathbf{p} - \mathbf{R}\mathbf{f}^{n-1}\right), \tag{6}$$

$$\mathbf{f}^n = H_K(\tilde{\mathbf{f}}^n), \qquad (7)$$

where $\mu$ is the step-size and $H_K(\tilde{\mathbf{f}}^n)$ is the non-linear operator that sets all but the $K$ largest (in magnitude) elements in $\tilde{\mathbf{f}}^n$ to zero.[15]

From this, one can find that IHT is a very simple algorithm, it can be applied to image reconstruction[22] and it has been proved that IHT can recover near-optimal solutions of the sparse and approximately sparse signals under certain conditions. [17,19,23]

## 2.2. Algorithm development

In the CT reconstruction, a two-dimensional digital reconstructed image can be expressed as $\mathbf{f} = (f_{i,j}) \in \mathbb{R}^I \times \mathbb{R}^J$, and to avoid ambiguity, we define

$$f_n = f_{i,j} \quad n = (i-1) \times J + j \quad N = I \times J, \qquad (8)$$

to meet the model equation (5), and we will use both the sign $f_n$ and $f_{i,j}$ next for convenience.

However, in the practical CT reconstruction, sparsity of images in terms of voxels $\mathbf{f}$ is generally not a widely applicable assumption, which means that the model (5) cannot be used directly. As mentioned above, it is often the case that images have sparse DGT $\nabla \mathbf{f}$, then we change the model (5) as

$$\mathbf{f} = \arg\min \|\mathbf{R}\mathbf{f} - \mathbf{p}\|_2^2 \quad s.t. \quad \|\nabla \mathbf{f}\|_0 \leq S, \qquad (9)$$

where $\|\nabla \mathbf{f}\|_0$ is the number of nonzero elements of $\nabla f_{i,j}$,

$$\nabla f_{i,j} = \sqrt{(f_{i,j} - f_{i+1,j})^2 + (f_{i,j} - f_{i,j+1})^2}. \qquad (10)$$

To solve the above optimization problem (9), we propose the projection onto convex sets (POCS) iterative hard thresholding (IHT-POCS) algorithm. The IHT-POCS has following two main steps:

Step 1: data projection iteration

In the typical IHT, a gradient descent step (6) is used to minimize the objective function; in the CT reconstruction here, we use the POCS to replace it. We use the algebraic reconstruction technique (ART) algorithm instead of (6) to update the reconstruction: $\tilde{\mathbf{f}}^k = ART(\mathbf{f}^{k-1})$, that is

*for* $m = 1, 2, \cdots, M$

$$\tilde{\mathbf{f}}^{k,0} = \mathbf{f}^{k-1},$$

$$\tilde{\mathbf{f}}^{k,m} = \tilde{\mathbf{f}}^{k,m-1} + \lambda^k \frac{p_m - \mathbf{R}_m \tilde{\mathbf{f}}^{k,m-1}}{\mathbf{R}_m \cdot \mathbf{R}_m^T} \cdot \mathbf{R}_m^T. \tag{11}$$

Here, $k$ is the main iteration number and $\mathbf{R}_m$ is the $m$ th row of measurement matrix $\mathbf{R}$.

Step 2: IHT algorithm

$$\mathbf{f}^k = \nabla^{-1}\left(H_S\left(\nabla \tilde{\mathbf{f}}^k\right)\right). \tag{12}$$

As mentioned above, DGT of (10) is not invertible. Inspired by [7], we use the $\nabla^{-1}$ as the pseudo-inverse transform of DGT.

Assume that $w$ is the $S$ th largest element of $\nabla \tilde{\mathbf{f}}^k$, according to (12), when $\nabla \tilde{f}_{i,j}^k < w$, we should adjust the value of $\tilde{f}_{i,j}^k, \tilde{f}_{i+1,j}^k$ and $\tilde{f}_{i,j+1}^k$ to make $\nabla f_{i,j}^k = 0$; when $\nabla \tilde{f}_{i,j}^k \geq w$, we should keep the value of $\tilde{f}_{i,j}^k$, $\tilde{f}_{i+1,j}^k$ and $\tilde{f}_{i,j+1}^k$ unchanged to make $\nabla f_{i,j}^k = \nabla \tilde{f}_{i,j}^k$. This non-linear pseudo-inverse transform is

$$f_{i,j}^k = \frac{2 f_{i,j}^{k,a} + f_{i,j}^{k,b} + f_{i,j}^{k,c}}{4}, \tag{13}$$

$$f_{i,j}^{k,a} = \begin{cases} \dfrac{\tilde{f}_{i,j}^k + \tilde{f}_{i+1,j}^k + \tilde{f}_{i,j+1}^k}{3} & \nabla \tilde{f}_{i,j}^k < w \\ \tilde{f}_{i,j}^k & \nabla \tilde{f}_{i,j}^k \geq w \end{cases}, \tag{14}$$

$$f_{i,j}^{k,b} = \begin{cases} \dfrac{\tilde{f}_{i-1,j}^k + \tilde{f}_{i,j}^k + \tilde{f}_{i-1,j+1}^k}{3} & \nabla \tilde{f}_{i-1,j}^k < w \\ \tilde{f}_{i,j}^k & \nabla \tilde{f}_{i-1,j}^k \geq w \end{cases}, \tag{15}$$

$$f_{i,j}^{k,c} = \begin{cases} \dfrac{\tilde{f}_{i,j-1}^k + \tilde{f}_{i,j}^k + \tilde{f}_{i+1,j-1}^k}{3} & \nabla \tilde{f}_{i,j-1}^k < w \\ \tilde{f}_{i,j}^k & \nabla \tilde{f}_{i,j-1}^k \geq w \end{cases}. \tag{16}$$

The flow chart of IHT-POCS is shown in table 1, where $N_{iter}$ is the max number of reconstruction iterations; $\lambda$ is the relaxation parameter that can be set to one or other values, for example, $\lambda = 1.0 \times 0.99^k$; $\varepsilon^k$ is the $\ell_2$-distance between two iterative images $\mathbf{f}^k$ and $\mathbf{f}^{k-1}$. When $\varepsilon^k < \varepsilon_0$, this exit criterion means that there is no longer any appreciate change in the iteration and the IHT-POCS algorithm runs into convergence.

Table 1. Implementation steps of IHT-POCS reconstruction.

| Algorithm of IHT-POCS |
| --- |
| **Initialization:** <br> Given $\mathbf{R}$, $\mathbf{p}$, $N_{iter}$, $\lambda$, $S$ and $\varepsilon_0$ <br> $\mathbf{f}^0 = \mathbf{0}$ <br> **Main iteration loop:** <br> $for \quad k=1,2,3,\cdots,N_{iter} \quad do$ <br>     **ART updating:** <br>     $\tilde{\mathbf{f}}^{k,0} = \mathbf{f}^{k-1}$ <br>     $for \quad m=1,2,3,\cdots,M \quad do$ <br> $$\tilde{\mathbf{f}}^{k,m} = \tilde{\mathbf{f}}^{k,m-1} + \lambda^k \frac{p_m - \mathbf{R}_m \tilde{\mathbf{f}}^{k,m-1}}{\mathbf{R}_m \cdot \mathbf{R}_m^T} \cdot \mathbf{R}_m^T$$ <br>     $end$ <br>     **Positivity constraint:** <br> $$\tilde{f}_{i,j}^k = \begin{cases} \tilde{f}_{i,j}^{k,M} & \tilde{f}_{i,j}^{k,M} \geq 0 \\ 0 & \tilde{f}_{i,j}^{k,M} < 0 \end{cases}$$ <br>     **Iterative hard thresholding:** <br>     Calculate $\nabla \tilde{\mathbf{f}}^k$, $w$ <br>     Perform the non-linear filter $\mathbf{f}^k = \nabla^{-1}\left(H_S\left(\nabla \tilde{\mathbf{f}}^k\right)\right)$ <br>     by using (13) (14) (15) (16) <br>     **Exit criterion:** <br>     $\varepsilon^k = \left\|\mathbf{f}^k - \mathbf{f}^{k-1}\right\|_2$ <br>     $if \quad \varepsilon^k < \varepsilon_0 \quad then$ <br>         $exit$ <br>     $end$ <br> $end$ |

## 3. Simulation

In this section, simulation is performed to demonstrate the proposed conclusions and evaluate the performance of the IHT-POCS algorithm. We apply four methods to reconstruct the images: (1) the classical ART method without using any sparsity regularization; (2) the ART iteration with TV regularization method (ART-TV) as in [1]; (3) the iterative soft thresholding with TV regularization method (IST-TV) as in [7]; (4) the proposed IHT-POCS method using the $\ell_0$-norm of the DGT image. For all the above methods, the parameter $\lambda^k$ in the ART iteration in (11) is set to be constant 1.

In the first study, we assume that the sparsity level $S$ of the reconstructed image is known, this can be obtained by FBP reconstruction of a similar image using full scan projections. Firstly, we use the well-known $128 \times 128$ modified Shepp-Logan phantom and set $S = 1081$. Figure 1 shows the few views image reconstructions based on different methods from 21 projections after 800 iterations; Figure 2 shows the IHT-POCS reconstructions after different numbers of iteration. It should be noted that the performance of ASD-POCS is not as good as expected here, this may be partly because of the parameters we selected in ASD-POCS failed to balance TV-gradient descent with POCS. After trying different choices of parameters, we pick out the one that provide the best performance. Comparing Fig. 1d with other reconstructed images, one can find that the proposed IHT-POCS algorithm can overcome the TV model's disadvantageous tendency of over smooth the low-contrast structures, and also it can improve quality of reconstruction.

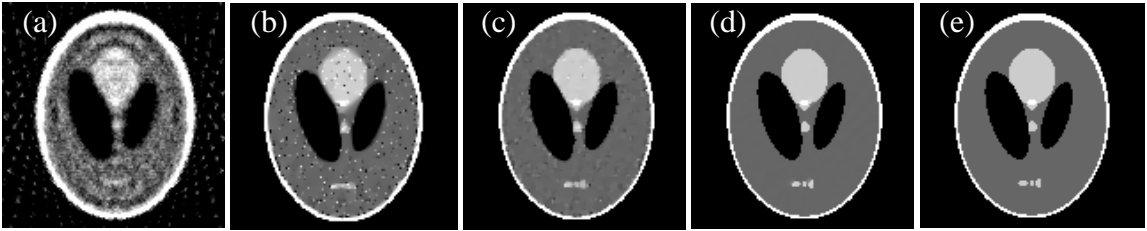

**Fig. 1.** Reconstructed images of a modified Shepp-Logan phantom from 21projections after 800 iterations. From (a) to (e), corresponding to the reconstructions using ART only, ART-TV, IST-TV, IHT-POCS and original phantom, respectively. The display window is [0.1, 0.35].

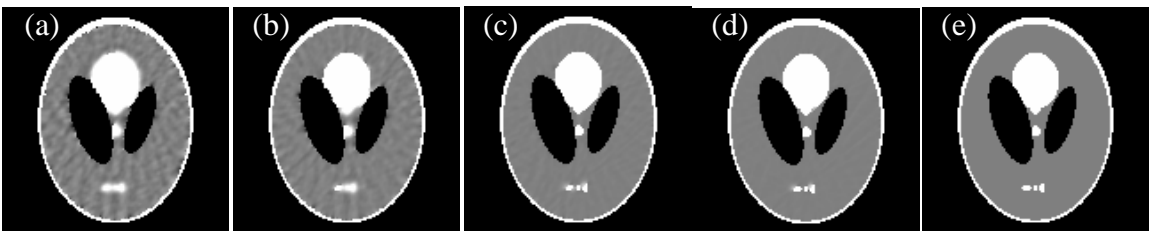

**Fig. 2.** The reconstructions based on IHT-POCS from 21 projections after different numbers of iterations. From (a) to (d), corresponding to the numbers of iterations are 100, 200, 400 and 800, respectively. The last image (e) is the ideal phantom. The display window is [0.15, 0.25].

Here we also use three evaluation parameters to evaluate the reconstructed image: the normalized mean square distance ($d$), the normalized absolute average distance ($r$),[24] and the peak signal to noise ratio ($PSNR$).

$$d = \left( \frac{\sum_{i=1}^{N}\sum_{j=1}^{N}(t_{i,j}-f_{i,j})^2}{\sum_{i=1}^{N}\sum_{j=1}^{N}(t_{i,j}-\bar{t})^2} \right)^{1/2}, \quad r = \frac{\sum_{i=1}^{N}\sum_{j=1}^{N}|t_{i,j}-f_{i,j}|}{\sum_{i=1}^{N}\sum_{j=1}^{N}|t_{i,j}|}, \quad (17)$$

where $t$ is the test standard image, $\bar{t}$ is the average value of $t$, $f$ is the reconstructed image. One can see that $d$ is sensitive to big errors, while $r$ is sensitive to whole errors.

Figure 3 shows the evaluation parameters of the reconstructions based on ART-TV, IST-TV, and IHT-POCS from 21 projections on each iteration. From this, one can see that the IHT-POCS algorithm can gain a much faster convergence and better reconstruction quality than the other algorithms.

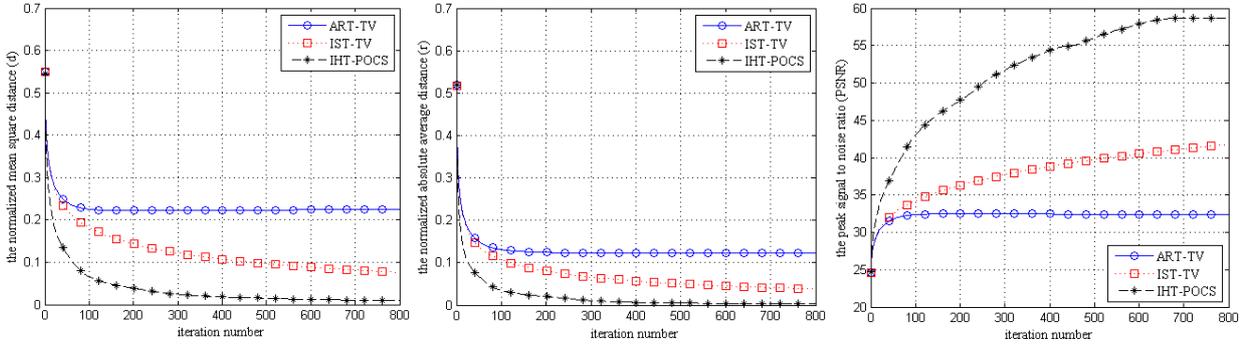

**Fig. 3.** The evaluation parameters of the reconstructions based on ART-TV, IST-TV, IHT-POCS: $d$, $r$ and $PSNR$.

Secondly, to challenge the proposed algorithm, we also repeat the above experiment in the $256\times 256$ 2D FORBILD head phantom.[25] Figure 4 shows the few views image reconstruction from 41 projections after 800 iterations. A similar simulation without IHT-POCS can be found in Figure 6 in [7], although it uses the simultaneous algebraic reconstruction technique (SART) instead of ART.

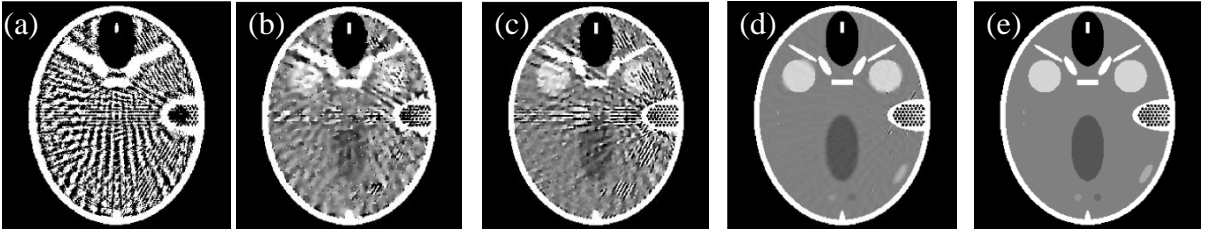

**Fig. 4.** Reconstructed images of FORBILD head phantom from 41 projections after 800 iterations. From (a) to (e), corresponding to the reconstructions using ART only, ART-TV, IST-TV, IHT-POCS and original phantom, respectively. The display window is [1.035, 1.065].

In the second study, the sparsity level $S$ of the reconstructed image is unknown. From (9), we can see that the data residual $\sigma(S) = \|\mathbf{P} - \mathbf{R}\mathbf{f}\|_2^2$ is depending on $S$. When $S$ is equal to $N$, then $\sigma(S)$ is close to 0. The overall trend is that $\sigma(S)$ varies inversely as $S$, but when $S$ is very close to the actual sparsity, $\sigma(S)$ reaches a local minimum; and once $S$ is smaller than the actual sparsity, then $\sigma(S)$ will increase suddenly. From the data error $\sigma(S)$ - sparsity $S$ chart, one can estimate the DGT image sparsity level. This is illustrated on Fig. 5, there is a sudden change at $S = 2^{11}$, which means that the real sparsity is between $2^{11}$ and $2^{10}$; then we can use the dichotomy searching method to make its scope further narrowed, and our several experiments indicate that it would be better that the estimation is loose and a little greater than the actual value. Here, the actual sparsity $S_* = 1081$ and $\sigma(S_*) = 0.4227$. If the sparsity level is unknown, we can estimate it first and then reconstruct the image by using the IHT-POCS algorithm based on the estimated $S$.

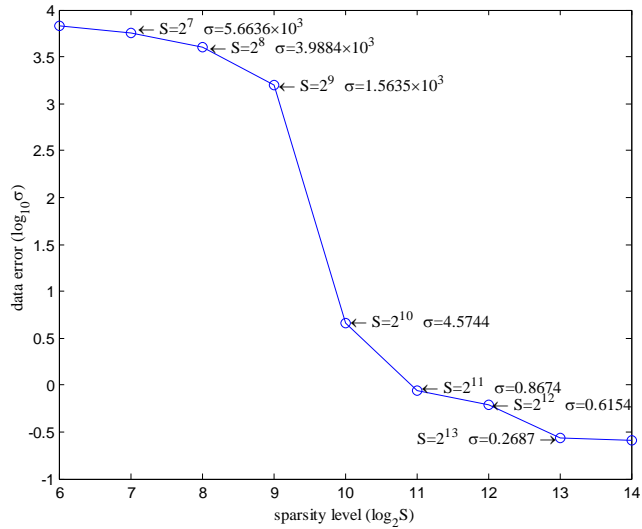

**Fig. 5.** Data errors at different sparsity level estimations, under the same conditions as in Fig. 1.

## 4. Conclusion

In this paper, we propose a new iterative algorithm to reconstruct the image from few views, which use the $\ell_0$-norm of the DGT as the regularization. The proposed algorithm uses a pseudo-inverse transform of DGT and adapts the IHT algorithm to address the challenges introduced by the $\ell_0$-norm DGT. The simulation shows that this algorithm can effectively improve the reconstruction quality. Meanwhile, as shown in the simulation, the sparse level can only be roughly estimated, so how to obtain the accurate estimation is still unresolved for us, which is also our next work. In the near future, we will evaluate the algorithm with physical as well as clinical datasets, and investigate other reconstruction methods to deal with the $\ell_0$-norm optimization. As the few views reconstruction has varies application demands in many fields, we believe that the proposed reconstruction algorithm is expected to have potential practical merits.